%% file: Vjets.tex
\documentclass{desyproc}
\usepackage{sidecap}
\usepackage{lineno}
\begin{document}

\title{$V$+jets Background and Systematic Uncertainties in Top Quark Analyses}


%

%
\author{{\slshape Stefanie Adomeit$^1$, Reinhild Yvonne Peters$^{2,3}$ for the ATLAS, CDF, CMS and D0 Collaborations}\\[1ex]
$^1$ Ludwig-Maximilians-Universit\"at M\"unchen, Fakult\"at f\"ur Physik, Schellingstra{\ss}e 4, 80799 M\"unchen, Germany\\
$^2$ The University of Manchester, Manchester, M13 9PL, UK \\
$^3$ DESY, Notkestra{\ss}e 85, 22607 Hamburg, Germany}

%

\contribID{xy}  
\confID{7095}
\desyproc{DESY-PROC-2013-XY}
\acronym{TOP2013}
\doi            

\maketitle
\begin{abstract}
 Vector boson production in association with jets is an important process to test perturbative quantum chromodynamics and also a background process in top quark analyses. Measurements on vector boson production 
 in association with light and heavy flavour jets are presented,
 performed by the D0 and CDF collaborations at the Tevatron as well as
 the ATLAS and CMS experiments at LHC. Techniques applied in top quark analyses to estimate the vector boson+jets background are also discussed.
\end{abstract}

\section{Introduction}
\input{intro.tex}

\section{$V$+jets Measurements}
\label{VJetsMeasurements}
\input{measurements.tex}

\section{$V$+jets Background Modelling in Top Quark Analyses}
\input{TopVJets.tex}
\section{Conclusion}
\input{conclusions.tex}

\section*{Acknowledgments}
We would like to thank our collaborators from ATLAS, CDF, CMS and D0 for their help in preparing the
presentation and this article, in particular Stefano Camarda for
sharing his expertise with us. We also thank the staffs at Fermilab, CERN and
collaborating institutions, and acknowledge the support from the Helmholtz association and BMBF.


\begin{footnotesize}

\end{footnotesize}


\end{document}

%% file: intro.tex
Besides searching for new physics, one of the main purposes of hadron collider physics is the precision measurement of processes as predicted by the standard model (SM) of particle physics, and the comparison with theory predictions. 
Important processes are $W$+jets and $Z$+jets (in short: $V$+jets) production. 
Understanding the $V$+jets production processes is important for several reasons. First, inclusive and differential - i.e. as a function of one or more variables - cross sections provide a crucial test of perturbative quantum chromodynamics (pQCD) calculations. Secondly, in many searches and measurements, for example in top quark physics, they constitute one of the major background processes. Finally, the comparison of $V$+jets measurements with predictions using Monte Carlo (MC) generators helps to improve the tuning or choice of appropriate generators for the modelling. 

In the following, measurements of $V$+jets production from the D0 and CDF experiments at the Tevatron proton-antiproton collider at Fermilab and from the ATLAS~\cite{atlas} and CMS~\cite{cms} experiments at the LHC at CERN are discussed. Furthermore, the methods to model $V$+jets background processes in top quark analyses are reviewed.  The different definitions of algorithms used for the identification of jets are discussed elsewhere~\cite{jets}. 

%% file: measurements.tex
The various experiments at the Tevatron and the LHC have released several different measurements of cross sections for $W$+jets and  $Z$+jets production in association with light and heavy flavour jets. The measurements are compared to a variety of predictions. 
These predictions can be classified into different categories:
\begin{itemize}
\item MC generators using leading-order (LO) matrix elements (MEs) plus HERWIG~\cite{herwig} or PYTHIA~\cite{pythia} for parton shower and hadronisation. These MC generators include ALPGEN~\cite{alpgen}, MADGRAPH~\cite{madgraph} and SHERPA~\cite{sherpa}.
\item MC generators using fixed-order next-to-leading-order (NLO) calculations. In particular, these include BlackHat+SHERPA~\cite{blackhat}, Rocket+MCFM~\cite{rocket,mcfm} and MCFM.
\item MC generators using fixed-order NLO calculations plus a generator (PYTHIA or HERWIG) for parton showering and hadronisation. This class includes for example POWHEG~\cite{powheg}, MC@NLO~\cite{mcatnlo_1}, MENLOPS~\cite{menlops} and MEPS@NLO\cite{mepsnlo_1,mepsnlo_2}.
\item Calculations using wide angle resummations: HEJ (high energy jet)~\cite{hej}.
\item Approximate next-to-next-to-leading-order (NNLO) calculations: LOOPSIM+MCFM~\cite{loopsim}.
\item NLO QCD calculations including NLO electroweak contributions. 
\end{itemize}

For most measurements, several of these predictions are directly compared to the shapes of the differential distributions after correcting for resolution effects, enabling to draw conclusions of the validity of the models for different variables. 
In the following, an overview of recent $W$+jets and $Z$+jets measurements from Tevatron and LHC are given. Details of the event selection and methods for unfolding are omitted in the following and can be found in the respective references of the analyses. 

\subsection{$W$+jets Measurements}
Various measurements of the $W$+jets process were performed at Tevatron and LHC, with different numbers of jets considered. In all measurements, the $W$ boson is required to decay into a charged lepton (usually electron or muon) and the associated neutrino. 
Using 320~pb$^{-1}$ of data, CDF performed measurements of the total and differential $W$+jets cross section, where 1 to $\geq$ 4 jets are considered~\cite{cdfwjets}. Recently, the D0 collaboration released a  measurement of  the inclusive and differential $W$+jets cross section, using events where the $W$ boson decays into an electron and the associated neutrino. With 3.7~fb$^{-1}$ of data, the differential cross section as function of about 40 different variables has been studied~\cite{d0wjets}.   These variables include jet and lepton energy and angular variables, dijet rapidity separations and opening angles, dijet azimuthal angular separations and the $W$ boson transverse momentum. Additionally, the number of jets is measured as function of the scalar sum of the transverse energies of the $W$~boson and the jets and as function of rapidity separations between the jets. The predictions of various calculations, compared to the measured distributions, vary more than the experimental 
uncertainty, enabling the 
usage of the measurements for improved modelling of $W$+jets. The ATLAS and CMS collaborations also performed measurements of $W$+jets, using 36~pb$^{-1}$ of 7~TeV LHC data~\cite{atlaswjets,cmswjets}, showing in general good agreement between prediction and measurement.

Besides $W$+jets processes with jets not distinguished according to their flavour, it is crucial to measure $W$+heavy flavour jet cross sections. These processes contribute an important background in several searches and measurements, in particular analyses where $b$-jet identification is applied. 

Measurements of $W$+$b(b)$~\footnote{In the following, $V+b(b)$ refers to $V+b$, $V+\bar{b}$, and $V+b\bar{b}$. The same applies for $V+c(c)$.} are interesting to test pQCD predictions in the presence of heavy quarks. A measurement of $W$+$b(b)$ from 1.9~fb$^{-1}$ of data by the CDF collaboration has been performed, using at least one identified $b$-jet~\cite{cdfwbjets}. The total cross section has been measured about three standard deviations higher than the NLO prediction. A recent analysis of $W$+$b(b)$ by the D0 collaboration, using 6.1~fb$^{-1}$ of data, has shown good agreement between the measured fiducial cross sections and predictions from SHERPA, MCFM and MADGRAPH~\cite{d0wbjets}. In this analysis, a lifetime based multivariate analysis technique has been used to distinguish the flavour contents of the $W$+jets samples. The ATLAS collaboration has studied $W$+$b(b)$ events with one identified $b$-jet on the full 7~TeV data sample of 4.6~fb$^{-1}$~\cite{atlaswbjets}. The flavour discrimination is 
done 
by a template fit of the neural network output distribution that is based on $b$-jet lifetime information. The 
fiducial cross section for one and two jets has been measured and compared to pQCD predictions, which show a good agreement for events with two jets, while for events with one jet the compatibility of measurement and prediction are only to the level of 1.5 standard deviations. In this analysis, also differential cross section measurements have been performed, for example as function of the $b$-jet transverse momentum ($p_T$). In this variable, the MCFM and ALPGEN predictions show a slight underestimation of the cross section for large $b$-jet $p_T$. Also using the full 7~TeV data sample of 5~fb$^{-1}$, the CMS collaboration has studied the $W$+$b(b)$ process with the $W$ boson decaying into a muon and associated neutrino~\cite{cmswbjets}. For this analysis, two identified $b$-jets are required. In order to distinguish events from $W$+$c\bar{c}$ and $W$+$b\bar{b}$, the sum of the invariant mass of the secondary vertex within each jet is used. The fiducial cross section shows good agreement with NLO 
predictions from MCFM. The studied kinematic distributions show also good agreement with MC predictions.

\begin{figure}[ht]
\centerline{\includegraphics[width=0.53\textwidth]{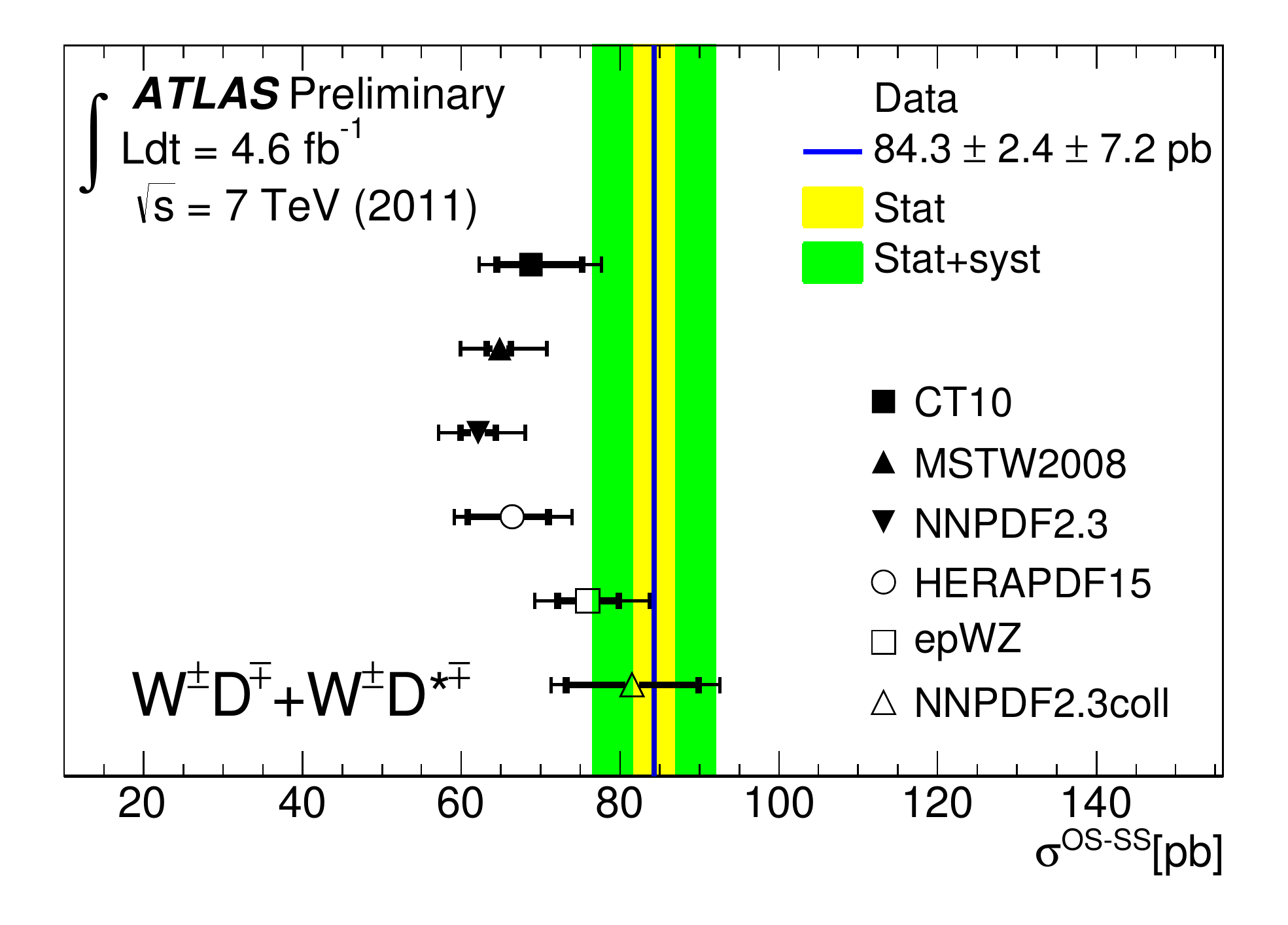}
\includegraphics[width=0.45\textwidth]{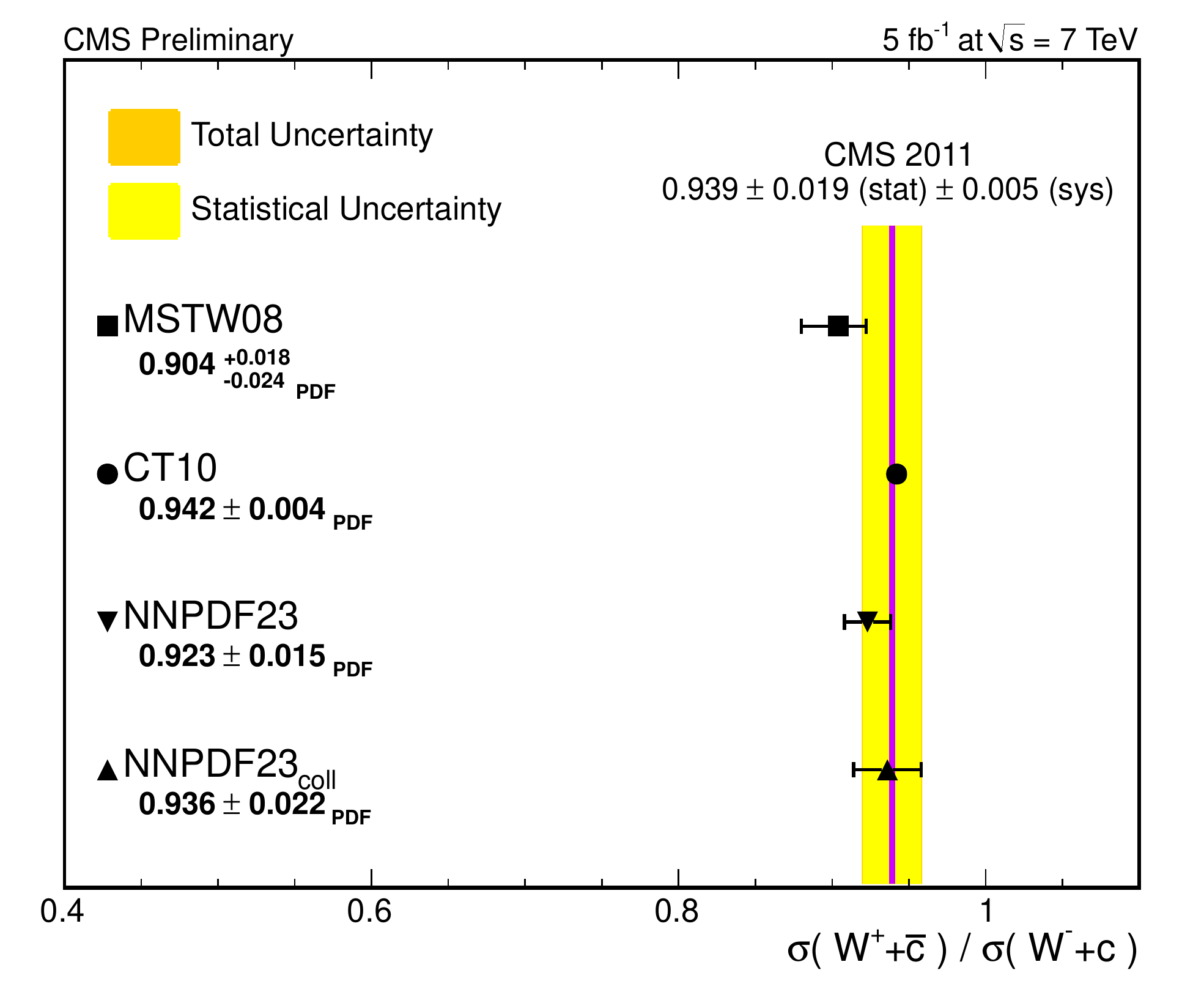}}
\caption{Measured cross sections $\sigma(W^{\pm}D^{\mp})+\sigma(W^{\pm}D^{*\mp})$~\cite{atlaswcjets} (left) and measured cross section ratio $\sigma(W^{+}+\bar{c})/\sigma(W^{-}+c)$~\cite{cmswcjets} (right)  compared to different PDF
predictions.}\label{fig:wc}
\end{figure}

The study of the $W$+$c(c)$ process is interesting for various reasons. In particular, the process $W$+$c$ can be used to probe the strange quark content of the proton and therefore its measurement is useful to distinguish different parton distribution functions (PDFs). Both the ATLAS and the CMS collaboration have performed $W$+$c(c)$ cross section measurements. The signature of these events is an opposite sign of the $W$ boson and a $D$ meson. At ATLAS, a measurement on 4.6~fb$^{-1}$ of 7~TeV data has been performed, where the $D$-meson is reconstructed from track information~\cite{atlaswcjets}. In this measurement, the cross section ratios $\sigma(W^{\pm}D^{\mp})/\sigma(W^{\pm})$ are measured inclusively and also differentially as function of the transverse momentum of the $D$-meson and the pseudorapidity of the lepton from the $W$ boson decay. In general, the results show good agreement with predictions, but it can also be observed that some of the PDF sets show a tension with the measurements. 
Figure~\ref{fig:wc} (left) shows the measured cross section sum $\sigma(W^{\pm}D^{\mp}) +\sigma(W^{\pm}D^{*\mp})$ compared to different PDF sets. PDF sets where the  $s$-quark sea is suppressed relative to the $d$-quark show worse agreement with the measurement than PDFs where this suppression is not included. 
Using 5.0~fb$^{-1}$ of 7~TeV data, CMS performed measurements of the total cross section of  $W$+$c$ and cross section ratios as well as measurements differentially as function of the pseudorapidity of the lepton from the $W$ boson
decay~\cite{cmswcjets}. In this analysis, $c$-jet candidates are identified using secondary vertex information. Figure~\ref{fig:wc} (right) shows the ratio $\sigma(W^{+}+\bar{c})/\sigma(W^{-}+c)$, compared to different predictions. A good agreement with predictions can be seen.

\subsection{$Z$+jets Measurements}
\begin{figure}[ht]
\centerline{\includegraphics[width=0.36\textwidth]{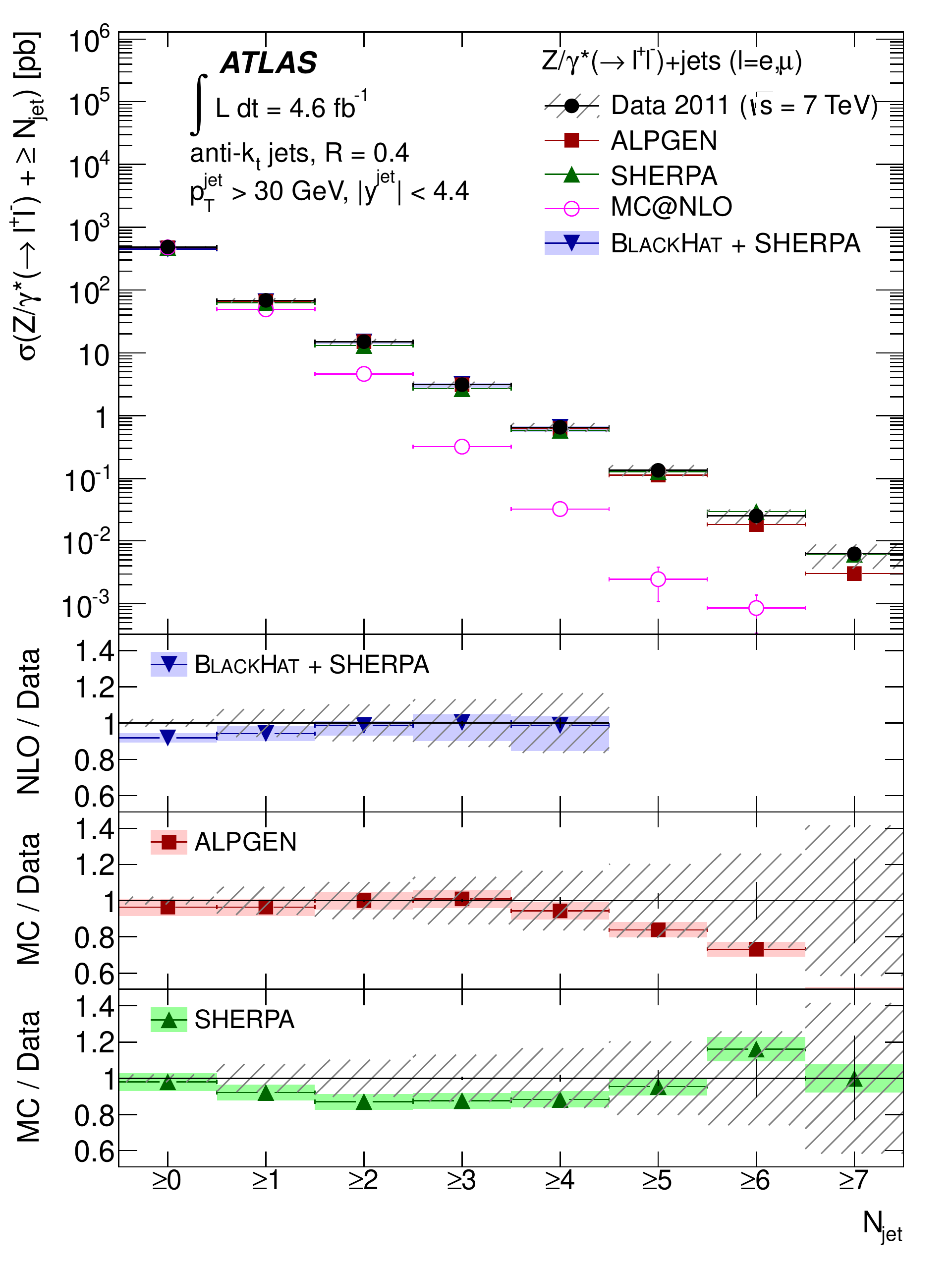}
\includegraphics[width=0.50\textwidth]{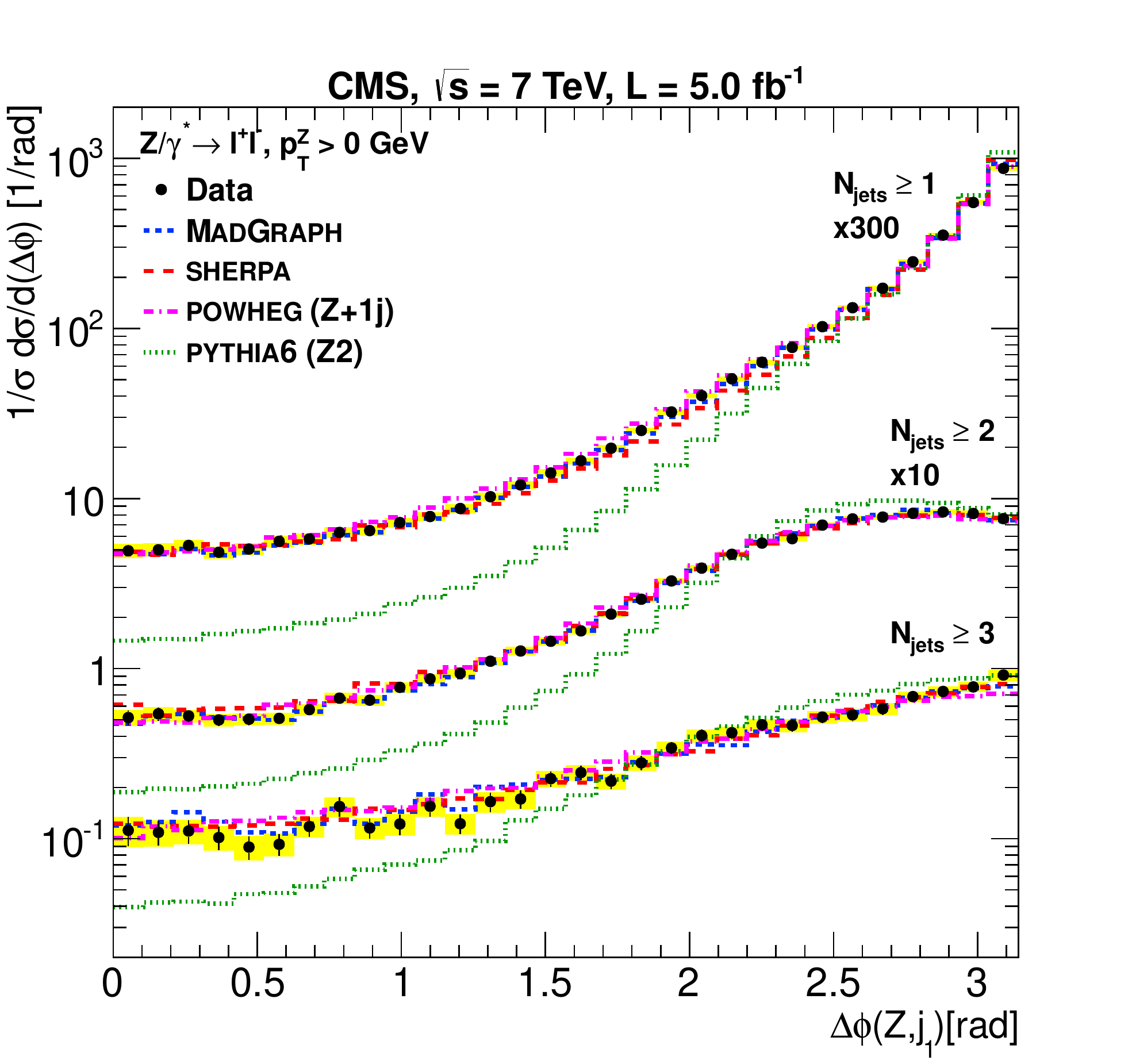}}
\caption{Measured cross section for $Z$+jets as a function of the inclusive jet multiplicity~\cite{atlaszjets} (left), and measured distributions of $\Delta \phi (Z, J_1)$ for different jet multiplicities~\cite{cmszjets} (right).}\label{fig:zjets}
\end{figure}

The experiments at Tevatron and LHC have also performed measurements of the $Z$+jets production~\footnote{$Z$+jets refers here to $Z/\gamma^{*}$+jets}, where the $Z$~boson is required to decay into a pair of oppositely charged leptons (usually electrons or muons). 
Using 1.0~fb$^{-1}$ of data, the D0 collaboration measured the $Z$+jets total cross section and differential cross section as function of the transverse momenta of the three leading jets~\cite{d0zjets}. Comparison of the measurements with LO and NLO pQCD predictions and different event generators shows good agreement. The CDF collaboration performed a $Z$+jets analysis using the full Tevatron Run~II data sample of 9.64~fb$^{-1}$~\cite{cdfzjets}, where the absolute and differential cross sections are measured, the latter as function of different variables. In general, good agreement between measurement and pQCD predictions could be observed. In a recent analysis by ATLAS, using the full 7~TeV data sample of 4.6~fb$^{-1}$, inclusive and differential $Z$+jets cross section measurements with up to seven jets have been performed~\cite{atlaszjets}. The differential studies are done as function of jet mulitplicities, jet transverse momenta, and angular distributions. In addition, distributions are studied after a 
modified selection 
optimised for vector boson fusion processes. Here, NLO pQCD predictions show a good description of the data, as do matrix element plus parton shower generators, while MC@NLO+HERWIG badly models the distribution as function of the number of jets and underestimates the cross section for large jet $p_T$.  Figure~\ref{fig:zjets} (left) shows the measured $Z$+jets cross section as function of the number of jets ($N_{\mathrm{jet}}$). The CMS collaboration has measured azimuthal correlations and event shapes for $Z$+jets processes, as well as $Z$+1 jet rapidity distributions in 5.0~fb$^{-1}$ of 7~TeV data~\cite{cmszjets}. These comparisons show good agreement between predictions and measurement, except for predictions from PYTHIA. Figure~\ref{fig:zjets} (right) shows the distribution of the azimuthal angle $\Delta \phi (Z, J_1)$ between the transverse momentum vectors of the $Z$ boson and the first leading $p_T$ jet for events with at least one, two or three selected jets.

With the same reasoning as for studies of the $W$+jets heavy flavour cross section, it is also important to measure $Z$+jets cross sections with heavy flavour jets. The D0 collaboration recently released a measurement of the ratio $\sigma(Z+b)/\sigma(Z+\mathrm{jets})$ based on 9.7~fb$^{-1}$ of data~\cite{d0zb}. The measurement was done inclusively and differentially as function of  the jet transverse momentum, jet pseudorapidity, $Z$ boson transverse momentum, and the azimuthal angle between the $Z$ boson and the closest jet for events with at least one identified $b$~jet. The inclusive measurement shows good agreement with NLO predictions, while none of the predictions can fully describe all studied variables. Based on 9.1~fb$^{-1}$ of data, the CDF collaboration measured the inclusive and differential $Z$+$b$ cross section, where the differential cross section is studied  as function of jet $p_T$ and rapidity~\cite{cdfzb}. The total cross section is found to be larger by about a factor of 1.6~compared to the 
prediction from ALPGEN, while good agreement with the 
NLO prediction from MCFM was observed.   At the LHC, the ATLAS collaboration has measured the  cross section for $b$-jet production in association with a $Z$ boson in 36~pb$^{-1}$ of 7~TeV data~\cite{atlaszb}, showing good agreement with NLO pQCD predictions. More recently, the CMS collaboration explored 5.0~fb$^{-1}$ of 7~TeV data, extracting the total $Z+b(b)$ cross section as well as the ratio of the cross section of a $Z$ boson produced in association with any number of $b$-jets
relative to those containing any number of jets. In addition, kinematic properties are compared to MC predictions using MADGRAPH, showing some deviations between prediction and data. 

The first measurement of the cross section ratios $\sigma(Z+c)/\sigma(Z+\mathrm{jets})$ and  $\sigma(Z+c)/\sigma(Z+b)$ has recently been performed on 9.7~fb$^{-1}$ of Tevatron data by the D0 collaboration~\cite{d0zc}. 
The cross section ratios are measured inclusively and differentially as function of jet and $Z$ boson transverse momenta. The inclusive measurements are not in agreement with pQCD predictions and predictions from different event generators. Furthermore, none of the predictions can fully describe the dependencies on all studied variables. It was found that an improved description of the distributions could be found by enhancing the $g \rightarrow c\bar{c}$ fraction in PYTHIA by an  empirical factor of 1.7.

\subsection{Summary of $V$+jets Measurements}
Both the Tevatron and the LHC experiments have performed a variety of $V$+jets measurements. These are in general in good agreement with NLO pQCD predictions and predictions from multi-leg generators, but also several variables have been identified that are not well modeled by current event generators. These studies are important input for future improvement of the choice and tuning of MC generators and to constrain PDF sets, as is possible with the $W$+$c$ jet measurement. The inclusive cross section measurements at LHC are in general limited by systematic uncertainties, in particular by uncertainties on the jet energy scale, while for differential measurements, especially for some parts of phase space, the results are still limited by the statistics of the data sample. Currently, the experiments work on further $V$+jets measurements, in particular on exploring the 8~TeV LHC data sample.  


%% file: TopVJets.tex
$V$+jets events are among the major background processes in measurements involving top quarks. $W$ boson production in association with heavy flavour jets can result in final states identical to those 
originating from the decay of a top-antitop quark pair in the lepton+jets channel as well as (t-channel) single top quark production. $Z$ boson + heavy flavour jets production together 
with ${\not\mathrel{E}}_T$ due to mismeasured objects leads to signatures similar to dileptonic top-antitop quark pair decays.

Both LHC and Tevatron experiments use LO matrix element (ME) generators (ALPGEN or MADGRAPH) for the modelling of $V$+jets background processes, 
interfaced to PYTHIA or HERWIG as parton shower (PS) MC. 
As heavy flavour jets can originate from both the ME and the PS, generating $V$+light jets and $V$+heavy flavour jets events separately can give rise to the same heavy flavour final states in the multiple samples. Techniques 
to remove this heavy flavour overlap need to be applied, based for instance on the opening angles between jets. 

Only few analyses use theoretical calculations to obtain the normalisation of $V$+jets background events. Uncertainties on the theoretical cross section predictions increase with increasing jet multiplicity, resulting in 
large uncertainties in jet multiplicity bins relevant for top quark analyses. Thus, the normalisation of $V$+jets background events is usually directly measured in the data while the shapes of $V$+jets
distributions are typically taken from MC simulation. $V$+jets measurement results as presented in Section~\ref{VJetsMeasurements} are, however, not directly used for this purpose so far as these cover phase space regions 
different from the ones used in top quark analyses. Top quark analyses hence apply their own dedicated techniques to measure the normalisation and, in many cases, the heavy flavour composition of 
$V$+jets background events. These techniques are discussed in Sections~\ref{WJetsBackground}~and~\ref{ZJetsBackground}.

\subsection{$W$+jets Background Determination Using Data}
\label{WJetsBackground}
Due to its signature $W$+jets events are most relevant as background process to lepton+jets $t\bar{t}$ and single top events. $W$+jets production 
together with a so-called fake lepton - originating from misidentified jets or leptons from semileptonic $c$/$b$ hadron decays - are among the background 
processes of dileptonic $t\bar{t}$ decays. Such fake lepton backgrounds are typically modelled using the so-called matrix method together with leptons fulfilling different 
categories of isolation criteria. A description on the determination of the fake lepton background can be found in~\cite{atlastTdlepXS}. In the following a selection of data-driven techniques  
applied by the two Tevatron and LHC experiments is summarised - with focus on the determination of the $W$+jets background normalisation and heavy flavour composition in lepton+jets 
$t\bar{t}$ analyses.  

\begin{SCfigure}[40][t!]

\includegraphics[width=0.36\textwidth]{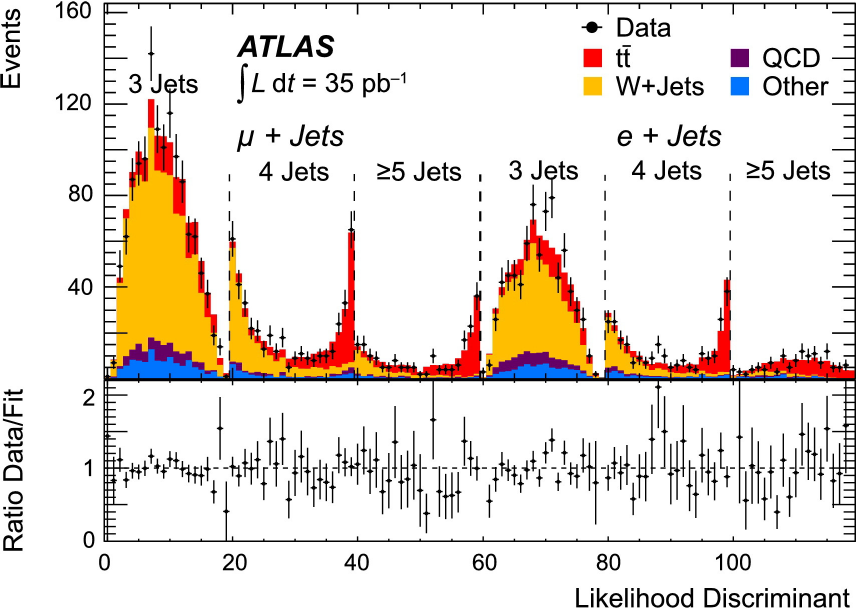} 
\includegraphics[width=0.34\textwidth]{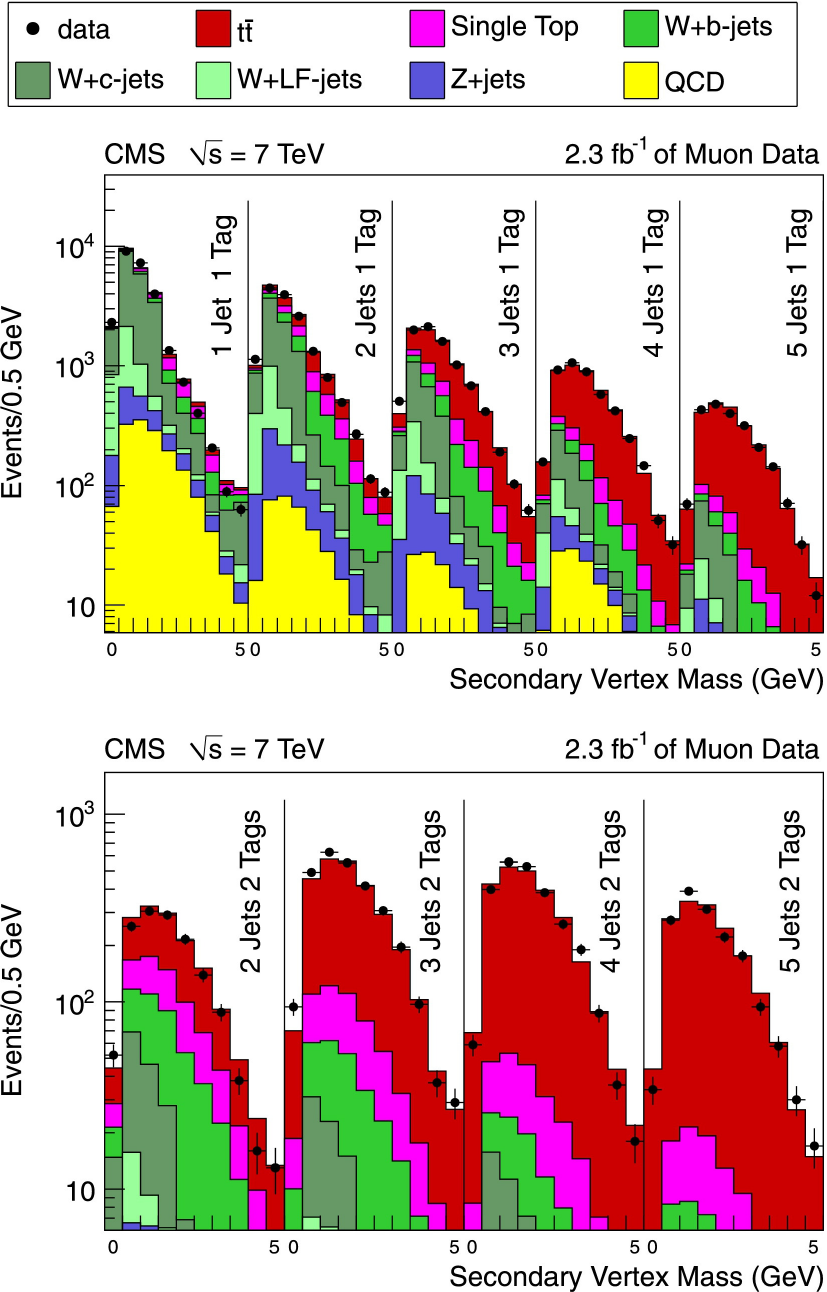}
\caption{Results of the combined fit to data in the ATLAS lepton+jets $t\bar{t}$ cross section measurement~\cite{atlastTlJetsXS}: The distribution of the likelihood discriminant for 
data is shown, superimposed on expectations for signal and backgrounds, scaled to the results of the fit (left). Results of the combined fit for the muon+jets channel for single $b$-tag and $\geq$2$b$-tag events in the CMS lepton+jets $t\bar{t}$ cross section measurement~\cite{cmstTlJetsXS} (right).}
\label{fig:cmsAtlastT}
\end{SCfigure}

To estimate the $W$+jets normalisation as well as its heavy flavour composition, the $W$ charge asymmetry technique is widely used in ATLAS. 
The $W$ charge asymmetry method~\cite{atlasCA} makes use of the larger number of $u$ valence quarks w.r.t. $d$ valence quarks in protons, resulting in 
an asymmetric  production rate of $W^{+}$+jets and $W^{-}$+jets events in proton-proton collisions. The total number of $W$+jets 
events can be measured in data according to
\begin{equation}
 N_{W^{+}}+N_{W^{-}}=\left( \frac{r_{MC}+1}{r_{MC}-1}\right)(D^{+}-D^{-}),
 \label{equ:1}
\end{equation}
where $N_{W^{+}}$ ($N_{W^{-}}$) is the number of $W^{+}$+jets ($W^{-}$+jets) events, $D^{+}$ ($D^{-}$) is the number of events containing a positively (negatively)
charged lepton after subtraction of non $W$+jets contributions and $r_{MC}=N(pp\rightarrow W^{+}$+jets)/$N(pp\rightarrow W^{-}$+jets), evaluated from MC. As the $W$ charge asymmetry is also 
sensitive to the heavy flavour composition of $W$+jets events, the relative fraction of $W$+$b\bar{b}$+jets, $W$+$c\bar{c}$+jets, 
$W$+$c$+jets and $W$+light jets events is extracted simultaneously with the overall normalisation in the $W$+jets dominated 2-jet bin and is then extrapolated to the bins with more than two jets.
As an alternative approach \cite{atlastTlJetsXS} measures the $W$+jets normalisation simultaneously with the $t\bar{t}$ lepton+jets cross section by means of likelihood discriminants, constructed from variables 
chosen for their signal vs. background discriminating power, see Figure~\ref{fig:cmsAtlastT} (left).

Similarly to the ATLAS lepton+jets $t\bar{t}$ cross section analysis, the W+jets normalisation and the $t\bar{t}$ cross section are extracted simultaneously in CMS~\cite{cmstTlJetsXS}. 
For this extraction a profile likelihood fit to the distribution of invariant masses of particles belonging to identified displaced vertices (secondary vertex mass, SVM) is performed as a function of 
jet and $b$-tag multiplicity. Due to 
the discriminating power of the SVM between heavy and light flavour jets, the normalisation is evaluated for each of the $W$+$b$, $W$+$c$, and $W$+light-flavour sub-samples, as shown in Figure~\ref{fig:cmsAtlastT} (right). Not only the W+jets yields but also the 
shapes of $W$+jets distributions are extracted from the data in the CMS single top t-channel cross section measurement~\cite{cmsSttXS}. This is done using a sideband region outside the top quark invariant 
mass window.

In the D0 lepton+jets $t\bar{t}$ cross section measurement~\cite{d0tTlJetsXS} the overall $W$+jets normalisation is extracted by subtracting all non $W$+jets events from the data, separately for each 
jet multiplicity bin. The relative contribution of $W$+heavy flavour jets (comprising $W$+$b\bar{b}$+jets and $W$+$c\bar{c}$+jets events), $W$+$c$+jets and $W$+light flavour jets events are determined using
NLO calculations and is verified in data using events with exactly one and two jets, split into subsamples with and without $b$-tagged jets. The $W$+jets normalisation is also measured simultaneously with the $t\bar{t}$ cross section, 
using a binned maximum likelihood fit for the predicted number of events in different jet and $b$-tag multiplicity bins. 

In the CDF lepton+jets $t\bar{t}$ cross section measurement~\cite{cdftTlJetsXS} the overall $W$+jets normalisation in pretag events is 
obtained by subtracting all non $W$+jets contributions from the data. The contribution of $W$+heavy flavour jets events 
to the $t\bar{t}$ signal region with at least one $b$-tagged jet is extracted using the overall $W$+jets normalisation in pretag events as well as 
MC simulation based predictions on the $W$+heavy flavour fractions together with the tagging efficiency of jets.
The fraction of $W$+heavy flavour jets is evaluated separately for $W$+$b\bar{b}$+jets, $W$+$c\bar{c}$+jets and $W$+$c$+jets events. Additonal correction factors to the MC based heavy flavour 
fractions are derived using a neural network fit to variables sensitive to jets matched to heavy and light flavour in 
dedicated control regions. Together with the mis-tag probability for light flavour jets - parameterised as a function of different 
jet variables - the flavour composition of $W$+jets events in the signal region can be determined. The heavy flavour contribution 
is also extracted together with the $t\bar{t}$ cross section via a simultaneous fit to a jet flavour discriminant across 
nine samples, defined by the number of jets and $b$-tags~\cite{cdftTlJetsXS_2}.

\subsection{$Z$+jets Background Determination Using Data}
\label{ZJetsBackground}
\begin{SCfigure}[1][ht]
\includegraphics[width=0.35\textwidth]{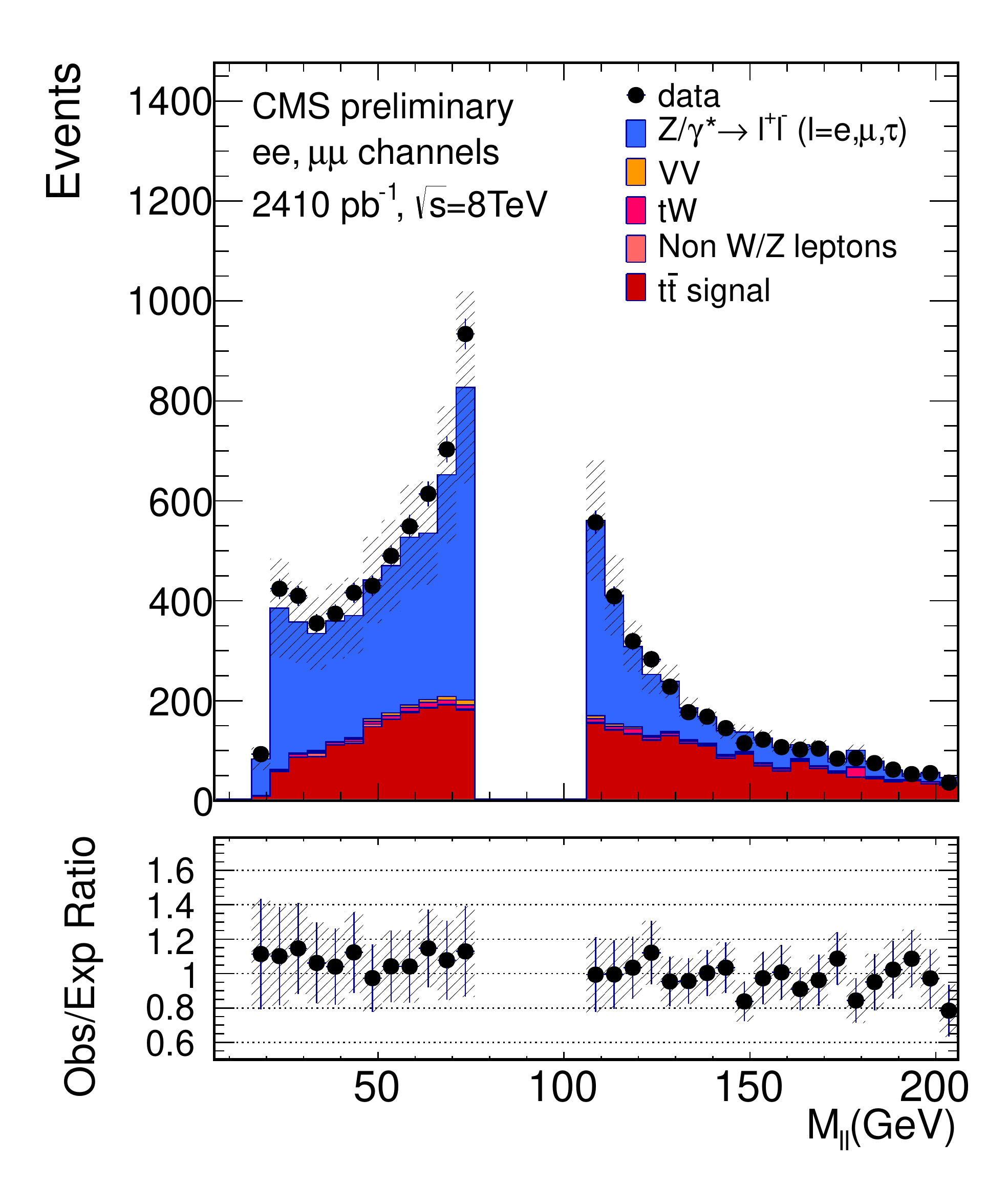}
\caption{Dilepton invariant mass distribution of the sum of the $e^+e^-$ and $\mu^+\mu^-$ channels in the CMS dilepton $t\bar{t}$ cross section measurement~\cite{cmstTldlepXS_2}. The gap in the $m_{ll}$ distribution reflects the requirement that removes dileptons 
from $Z$ decays. This gap defines the control region used for the estimate of the $Z$+jets background normalisation.}\label{fig:cmsDL}
\end{SCfigure}

$Z$+jets background processes in top quark analyses are most relevant to the dileptonic $t\bar{t}$ decay channel. To mimic 
the dileptonic $t\bar{t}$ final state, $Z$+jets events are required to have additional ${\not\mathrel{E}}_T$ from mismeasured objects. 
As such mismeasuring effects are difficult to model in MC simulation, the normalisation of $Z$+jets events is usually 
extracted from dedicated $Z$+jets enriched control regions in the data. These control regions are most commonly defined via 
the dilepton invariant mass $m_{ll}$, where the control region usually comprises events fulfilling a $Z$-mass window cut of 
$|m_{ll}-m_{Z}|<$10~GeV, see Figure~\ref{fig:cmsDL}. The $Z$+jets normalisation within the control region is determined in data and is extrapolated to 
the $t\bar{t}$ signal region ($|m_{ll}-m_{Z}|>$10~GeV) by means of scaling factors extracted from MC simulation~\cite{atlastTdlepXS,cdftTldlepXS}. The CMS dilepton 
$t\bar{t}$ cross section measurement~\cite{cmstTldlepXS} uses additional control regions to evaluate corrections to the MC based scaling factors 
using data. The ATLAS top polarisation measurement~\cite{atlastTlPol}
derives correction factors to the MC $Z$+jets normalisation as a function of ${\not\mathrel{E}}_T$, allowing to account for possible mismodelling of the 
${\not\mathrel{E}}_T$ distribution in MC simulation.

\subsection{$V$+jets Background and Systematic Uncertainties}
Thanks to the data-driven techniques, systematic uncertainties on the $V$+jets background normalisation are usually small. As an example, the $W$ charge asymmetry technique, as applied by 
the ATLAS collaboration,  results in correction factors to the $W$+jets normalisation of 0.83$\pm$0.14 and 0.94$^{+0.16}_{-0.14}$ in the elctron and muon+jets $t\bar{t}$ channel, respectively. The numbers refer
to events passing all signal selection criteria as outlined in ~\cite{atlasCA}, including 
the presence of $b$-tagged jets. Uncertainties on the MC modelling are usually accounted for by variations of the renormalisation and factorisation scales and, in 
 some cases, additional variations of generator internal cuts.

%% file: conclusions.tex
A variety of $V$+jets measurements has been performed by the D0 and CDF experiments at Tevatron and the ATLAS and CMS experiments at LHC. 
These measurements are an important input for the future improvement of MC generators. Most top quark analyses at both the Tevatron and LHC experiments use MC generators 
to model the shapes of the $V$+jets background distributions while different data-driven techniques are applied to obtain the normalisation as well as the heavy flavour composition in $V$+jets 
events. Thanks to these techniques the uncertainties on the $V$+jets background modelling are usually small in top quark analyses. Further improvement on the $V$+jets background modelling may be achieved 
when using the $V$+jets measurement results directly in top quark analyses, which, due to the different phase space regions covered by $V$+jets and top quark analyses, is currently not the case.